\documentclass[12pt]{article}

\usepackage{epsfig}
\usepackage{cite}
\usepackage{amssymb}
\begin{document}
\title{\Large Molecular simulation analysis of structural variations
in lipoplexes} \author{Oded Farago$^{1,2}$ and Niels
Gr{\o}nbech-Jensen$^{3,4}$ \\ 
$^1$Department of Biomedical Engineering and $^2$Ilse Katz Institute for Nanoscale\\ 
Science and Technology, Ben Gurion University, Be'er Sheva 84105, Israel.\\
$^3$Department of Applied Science, University of California, Davis,\\ 
California 95616, USA.\\ 
$^4$The Niels Bohr International Academy, The Niels Bohr Institute,\\ 
Blegdamsvej 17, 2100 Copenhagen, Denmark.} 
\maketitle
\begin{abstract}
We use a coarse-grained molecular model to study the self-assembly
process of complexes of cationic and neutral lipids with DNA molecules
(``lipoplexes'') - a promising nonviral carrier of DNA for gene
therapy. We identify the resulting structures through direct
visualization of the molecular arrangements and through calculations
of the corresponding scattering plots. The latter approach provides a
means for comparison with published data from X-ray scattering
experiments. Consistent with experimental results, we find that upon
increasing the stiffness of the lipid material, the system tends to
form lamellar structures. Two characteristic distances can be
extracted from the scattering plots of lamellar complexes - the
lamellar (interlayer) spacing and the DNA-spacing within each
layer. We find a remarkable agreement between the computed values of
these two quantities and the experimental data [J. O. R\"{a}dler,
I. Koltover, T. Salditt and C. R.  Safinya, {\em Science}\/, 1997,
{\bf 275} 810-814] over the entire range of mole fractions of charged
lipids (CLs) studied experimentally. A visual inspection of the
simulated systems reveals that, for very high fractions of CLs,
disordered structures consisting of DNA molecules bound to small
membrane fragments are spontaneously formed. The diffraction plots of
these non-lamellar disordered complexes appear very similar to that of
the lamellar structure, which makes the interpretation of the X-ray
data ambiguous. The loss of lamellar order may be the origin of the
observed increase in the efficiency of lipoplexes as gene delivery
vectors at high charge densities.
\end{abstract}
\newpage

When DNA molecules are mixed with neutral and cationic lipids (CLs) in
an aqueous environment, they spontaneously aggregate to form
macromolecular complexes called ``lipoplexes''. These complexes have
attracted much attention over the past two decades because of their
potential use as nonviral transfection vectors in gene therapy
\cite{general1,general2,general3,general4,general5}. Transfection is a
two-stage process involving adsorption and entry (via endocytosis) of
the lipoplex into the cell, followed by the release of the DNA to the
cytoplasm and delivery to the nucleus, which makes the DNA available
for expression
\cite{transfection1,transfection2,transfection3}. CL-DNA complexes
exhibit low toxicity and nonimmunogenicity, but their transfection
efficiency (TE) remains low compared to that of viral vectors
\cite{transfection1,transfection4}. This has spurred an intense
research activity aimed at enhancing TE. Recognizing that the
structure of CL-DNA complexes may strongly influence their function
and TE, much of the effort in theoretical and experimental studies has
been devoted to understanding the mechanisms governing complex
formation, structure, and phase behavior.  X-ray diffraction
experiments have revealed that CL-DNA complexes exist in a variety of
mesoscopic structures. These structures include: (i) a multilamellar
phase where DNA monolayers are intercalated between lipid bilayers
($L_{\alpha}^C$) \cite{safinya_science1}, and (ii) the inverted
hexagonal phase with DNA encapsulated within monolayers tubes and
arranged on a two-dimensional hexagonal lattice ($H_{II}^C$)
\cite{safinya_science2}.

One of the major advantages of lipoplexes over viral capsids is their
ease of preparation and their almost unlimited DNA-carrying capacity,
which stem from the fact that the vector is formed by spontaneous
self-assembly in aqueous solutions. The electrostatic attraction
between the anionic DNA and the CLs along with the entropic gain
associated with the release of tightly bound counterions from the CLs
and DNA are the driving forces for the formation of a complex.  In a
recent publication, we reported on coarse-grained (CG) simulations of
self-assembly of CL-DNA complexes \cite{farago_jensen_jacs}. We
demonstrated, in agreement with previous theoretical studies and X-ray
scattering experiments
\cite{safinya_science1,safinya_science2,harries_may_benshaul,harries_may_benshaul1},
that rigid membranes tend to form lamellar complexes. For soft
membranes, the preferred geometry is that of the inverted hexagonal
phase. Our simulations also revealed that the phase diagram of the
CL-DNA complexes is quite rich and includes, in addition to the
lamellar and inverted hexagonal complexes, several other disordered
structures with distinct configurational characteristics. We also
found a new ordered phase, which has thus far not been observed
experimentally, where DNA rods and cylindrical micelles form a 2D
square lattice analogous to the 3D cubic NaCl-type structure. Our
analysis of the computed self-assembled structures was based on
simulation images and on the calculation of the Fourier transform of
the DNA positions \cite{farago_jensen_jacs}. The Fourier transform
provides a quantitative measure for how the simulated structures would
appear in x-ray scattering experiments. To achieve a better comparison
with the experimental data, one should also consider the contribution
of the lipids to the scattered intensity. By plotting the separate
contributions of each component (something which cannot be done
experimentally), one can dissect the information displayed in the
scattering plots.

In this paper we analyze the scattering intensity plots (the square of
the Fourier transform averaged of all angles
\cite{farago_jensen_jacs}) of the lamellar complexes, which are formed
in our simulations when the DNAs are mixed with stiff lipid
material. The technical details of the model and the simulations have
been presented in ref.~\cite{farago_jensen_jacs}. In short, the model
is based on the Noguchi-Takasu implicit solvent CG membrane model
\cite{noguchi_takasu} in which each lipid of length $l_{\rm LIP}$ is
represented by a linear rigid molecule \cite{niels_erratum} consisting
of three beads of diameter $\sigma=l_{\rm LIP}/3=6.25$\AA, one of
which is hydrophilic and the other two are hydrophobic. The CLs are
modeled by associating the hydrophilic bead with a positive unit point
charge, while the DNA molecules are modeled as infinitely long
parallel rigid rods of diameter $D_{\rm DNA}=4\sigma=25$\AA~with
uniform charge density corresponding to -1.7e/\AA. The molecules
interact via three types of interactions: (i) Unscreened electrostatic
interactions which are calculated using the Lekner summation method
\cite{lekner1,lekner2}. (ii) Short-range repulsive (``hard core'')
potential. The bead-bead pair potential $U_{\rm rep,bb}$ is given by
Eq.~(4) in ref.~\cite{noguchi_takasu}, and the bead-DNA potential
$U_{\rm rep, bD}(r)= U_{\rm rep,bb}(r-1.5\sigma)$. Since the DNA rods
are strongly repelled from each other by electrostatic forces, there
was no need to introduce an additional $U_{\rm rep, DD}$. (iii) The
Noguchi-Takasu hydrophobic interaction potential, given by
Eqs.~(5)-(6) in ref.~\cite{noguchi_takasu}. The lipids and the DNA
rods are initially randomly distributed within a given volume in
the simulation box and, through Molecular Dynamics (MD) simulations at
constant temperature, we follow the evolution of the complexes under
different conditions. As the objective of the simulations is to
attain self-assembled structures representative of equilibrium, we
simulate tens of millions of time steps such that at least half of the
total simulated time does not change the characteristics of the
visible structures. We have also verified that, while the details of
the shown structures do depend on initial conditions, the significant
characteristics, such as the peaks in the resulting scattering
intensities, are well defined.
 
We study isoelectric complexes where the total charges on
DNA and the CLs neutralize each other, with no added counterions.  The
structure of the complex is determined as a function of two
parameters: (i) the fraction of CLs, $\phi_c$, which can be varied by
adding different amounts of neutral lipids (NLs), and (ii) the bending
modulus, $\kappa_s$, which is the prefactor of the bending energy term
introduced in the later version of the Noguchi-Takasu model
\cite{noguchi_takasu_2003} to control the stiffness of the simulated
membranes (see Eq.~(7) in ref.~\cite{farago_jensen_jacs}).

\begin{figure}[t]
\hspace{0cm}\includegraphics[width=16cm]{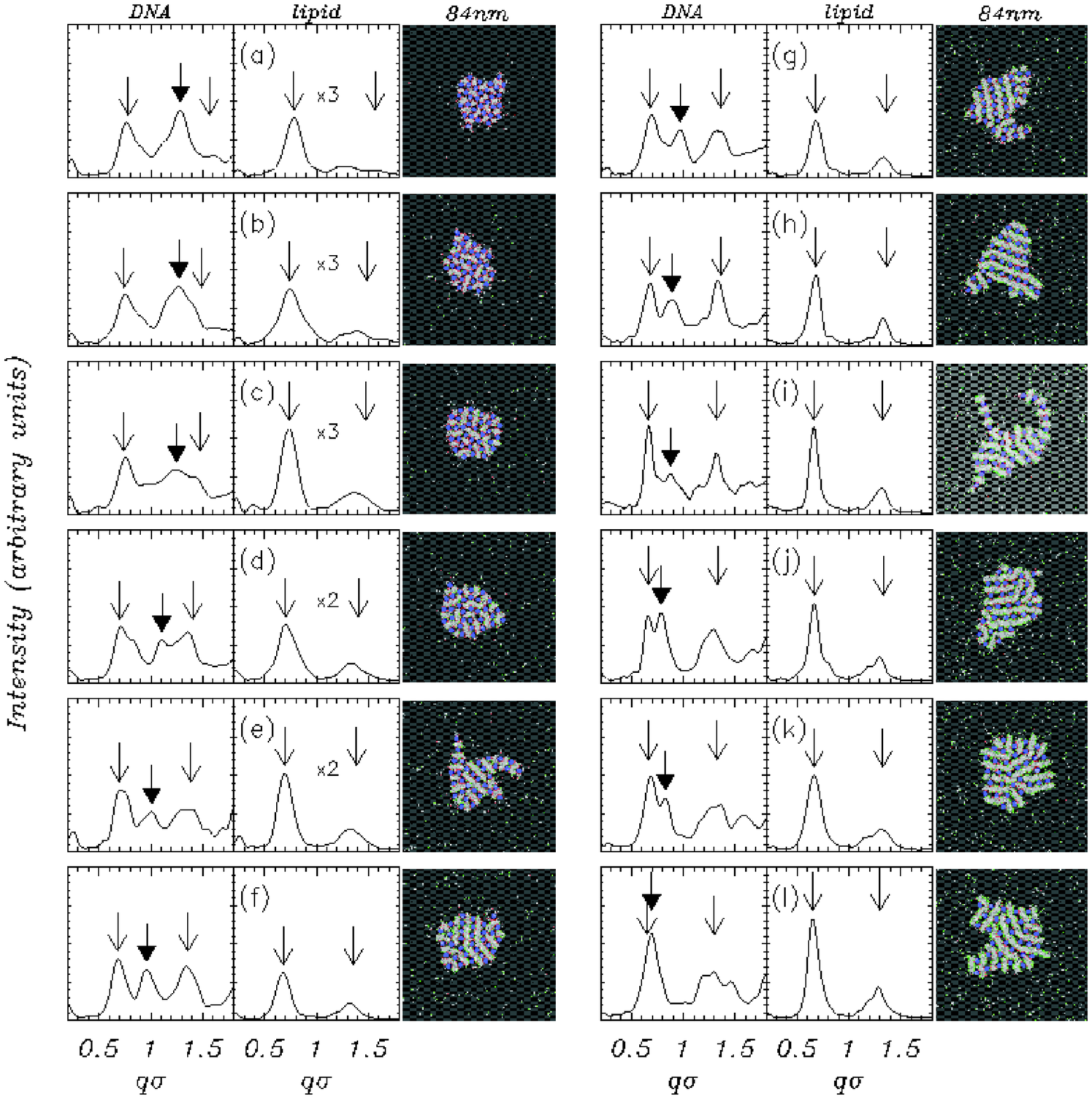}
\caption{Self-assembled complex structures consisting of 32 DNA rods
mixed with 800 CLs. The amount of NLs varies from 0 ($\phi_c=1$) at
(a) to 2200 ($\phi_c=4/15$) at (l). Each structure is initiated in a
random molecular configuration, and has evolved for $10-50\times 10^6$
MD time steps. The structures are viewed along the DNA axes. Color
coding: grey - hydrophobic lipid beads, red - charged hydrophilic
heads, green - neutral hydrophilic heads, and blue - DNA rods. For
each configurations, the scattering intensities of the DNA rods and
the lipids are also plotted. The open arrows indicate the position of
the lipid peaks at $q_{\rm LAM}$ and $2q_{\rm LAM}$. The solid arrow
indicate the position of the DNA in-plane correlation peak at $q_{\rm
DNA}$.  }
\label{fig:giant}
\end{figure}

\begin{figure}[t]
\hspace{0cm}\includegraphics[width=16cm]{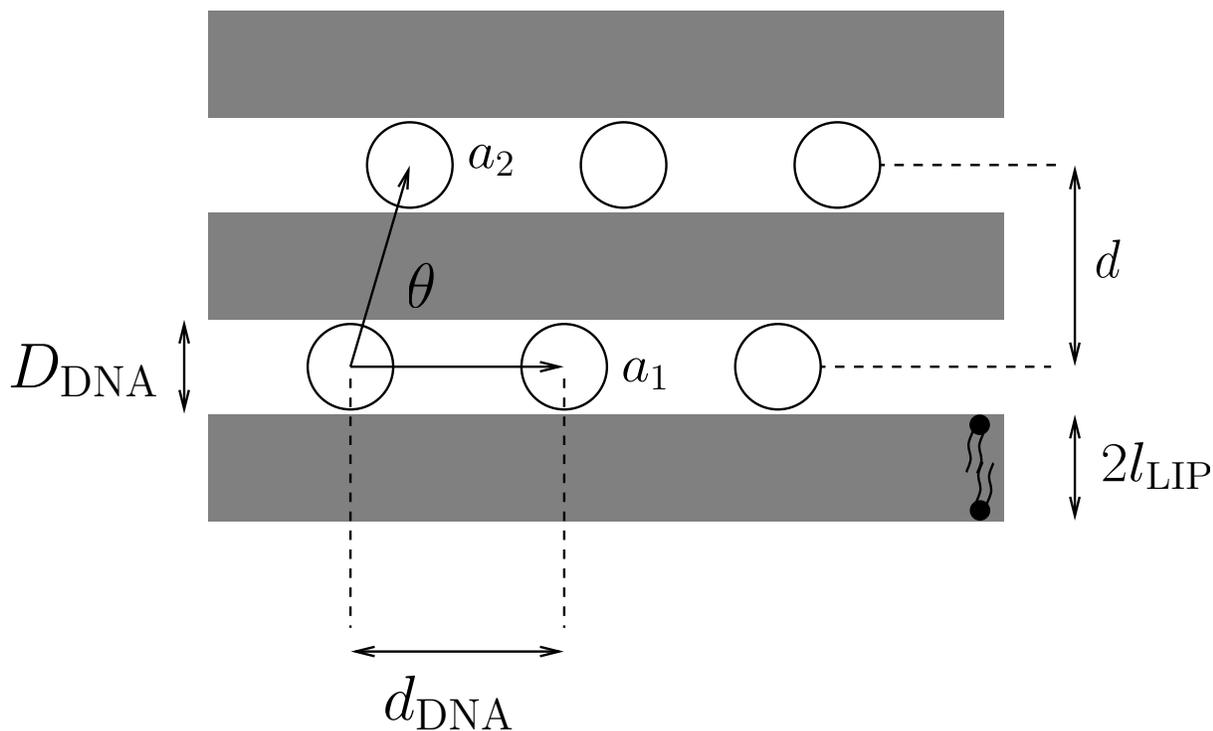}
\caption{Sketch of an idealized lamellar complex.}
\label{fig:ideal}
\end{figure}

Fig.~\ref{fig:giant} shows the diffraction patterns of stiff complexes
($\kappa_s=10$) with different values of $\phi_c$ ranging from
$\phi_c=1$ (a) to $\phi_c=4/15$ (l). Following the approach
outlined in ref.~\cite{farago_jensen_jacs}, we calculate diffraction
patterns from the two-dimensional (2D) Fourier transformation
\begin{eqnarray}
{\cal F}(\bar{q}) & = & \sum_{j=1}^Nw_j\exp(i\bar{r}_j\bar{q}) \; ,
\label{eq:Fourier}
\end{eqnarray}
where $\bar{r}_j$ represents the 2D coordinate of the DNA rod or the
2D coordinate of a bead's center of mass in the plane perpendicular to
the DNA axis, $\bar{q}$ is the reciprocal vector, and $w_j$ represents
the electron density of the $j$th particle relative to bulk water.
For each value of $\phi_c$, we show a triplet of figures consisting of
(left) the simulated scattered intensity from the DNA rods (with
$N=N_{DNA}$ in Eq.~(\ref{eq:Fourier})), (middle) the simulated scattered
intensity from the lipids ($N=3N_{\rm LIP}$), and (right) the
self-assembled structure of the complex. The displayed scattering
intensities are $I(q)\sim\langle|{\cal F}(\bar{q})|^2\rangle_\theta$, where
$\bar{q}=q\exp(i\theta)$ and $\langle\dots\rangle_\theta$ denote the
average over all angels.
All the DNA scattering plots are drawn on the same scale. The lipid
scattering plots are also drawn on the same scale, except for (a)-(e)
which are multiplied by the factor indicated on the corresponding
plot. The relative scattering intensity of the lipids and the DNA
depends on their electron densities, as well as on $\phi_c$. Using
reasonable values for the electron densities \cite{raviv1,raviv2}, we
find that the scale of the DNA intensities is two order of magnitude
larger than that of the lipids. Nevertheless, the scattering plots of
the lipids exhibit two well identified peaks located at $q_{\rm LAM}$
and $2q_{\rm LAM}$. These peaks are commonly associated with the
lamellar structure. From the position of the first lamellar peak, one
can extract the inter-layer lamellar spacing $d$ through $q_{\rm
LAM}=2\pi/d$. The DNA plots generally exhibit three peaks, two of
which coincide with the lamellar peaks from the lipid scattering
plots, and one which is commonly referred to as the ``DNA peak''. The
position of the latter at $q_{\rm DNA}$ provides information about the
DNA spacing within each layer of the lamellar structure, $d_{\rm
DNA}$. It is assumed that $q_{\rm DNA}=2\pi/d_{\rm DNA}$. Notice that
the scattering from the DNAs includes the information about the
lamellar spacing observed also in the scattering intensity from the
lipids. This can be easily understood by considering the idealized
lamellar structure sketched in fig.~\ref{fig:ideal}. In this
structure, the DNA rods form an oblique lattice with lattice vectors
equal to $a_1=d_{\rm DNA}$ and $a_2=d/\sin\theta$. The reciprocal
lattice is also oblique with the same angle $\theta$ between the
lattice vectors $b_1=2\pi/d$ and $b_2=2\pi/d_{\rm DNA}$. The peaks
which can be seen in the scattering plots of the DNAs correspond to
the reciprocal lattice vectors: $b_1$, $2b_1$, and $b_2$. These are
also the peaks which are usually observed in actual scattering
experiments \cite{safinya_science1}. For idealized lamellar
structures, the scattering intensity has peaks at other wavevectors
$q$ which correspond to linear combinations of integer-multiples of
$b_1$ and $b_2$. The positions of these peaks are $\theta$-dependent
\cite{raviv_paper}. In non-idealized complexes, like the ones in our
simulations, these peaks are usually very small and are hard to be
detected in the scattering plots.

\begin{figure}[t]
\hspace{0cm}\includegraphics[width=16cm]{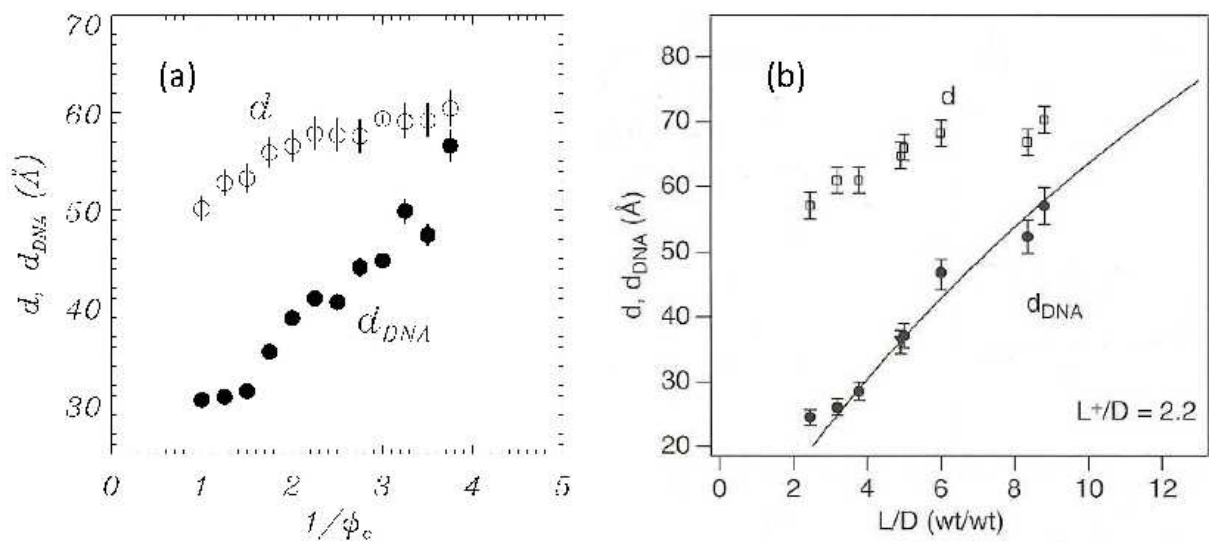}
\caption{(a) The inter-layer lamellar spacing $d$ and the DNA-spacing
$d_{\rm DNA}$ vs.~$1/\phi_c$, as computed from the peaks indicated by
arrows in the scattering plots shown in fig.~{\protect
\ref{fig:giant}}. (b) The same quantities derived from synchrotron
X-ray scattering data reported in ref.~{\protect
\cite{safinya_science1}}. $1/\phi_c=1$ in (a) corresponds to $L/D=2.2$
in (b). (b) has been adopted from {\protect
\cite{safinya_science1,safinya_currop}} with permission.}
\label{fig:companion}
\end{figure}

Two open arrows are drawn in each of the lipid scattering plots
included in fig.~\ref{fig:giant}. The first one indicates the position
of the larger lamellar peak which is located at $q_{\rm
LAM}=2\pi/d$. The second one denotes the wavevector $2q_{\rm LAM}$,
where we indeed find the second lamellar peak. These two arrows are
copied into the corresponding DNA scattering plot, verifying that
these peaks are also reproduced by the DNA ordering as explained
above. Notice that $q_{\rm LAM}$ is only weakly dependent on
$\phi_c$. In contrast, the position of the third peak, which is
indicated by the solid arrow, varies noticeably with $\phi_c$. This
peak is located at $q_{\rm DNA}=2\pi/d_{\rm DNA}$, and the variations
in its position reflect the decrease in the DNA spacing with
increasing $\phi_c$.  From the computed scattering plots shown in
fig.~\ref{fig:giant}, we can extract $d$ and $d_{\rm DNA}$ as a
function of $\phi_c$. Our results are summarized in
fig.~\ref{fig:companion} (a). The visual impression of this plot is
that the results are in overall agreement with the idealized geometry
of a lamellar structure shown in fig.~\ref{fig:ideal}. Specifically,
the lamellar spacing is well approximated by the sum of twice the
length of the lipids and the diameter of the DNA rods,
$d\approx(2l_{\rm LIP}+D_{\rm DNA})\approx
10\sigma=62.5${\AA}. Similarly, the characteristic distance $d_{\rm
DNA}$ between DNA rods in each layer approximates a linear
relationship with $1/\phi_c$
\cite{safinya_science1,oded_prl,oded_bpj}. This relationship can be
derived from the simple geometric consideration that equally spaced
DNA rods fill the surface area made available by the lipid bilayer
material for any given charge density. There are, however, noticeable
deviations from this idealized picture, especially at large
$\phi_c$. The lamellar spacing decreases below the ideal value, while
the DNA spacing attains values higher than predicted by the linear
relationship. The origin of these discrepancies becomes clear by
inspection of the corresponding structures shown in
fig.~\ref{fig:giant}, where we observe that the long range lamellar
order is lost at high $\phi_c$ in favor of a disordered arrangement of
smaller DNA-bilayer fragments. The transition is consistent with
previously described membrane rupture at high charge densities
resulting from the electrostatic stresses that develop in the complex
\cite{oded_prl,oded_bpj}. The highly charged cationic membrane
fragments strongly associate with the DNA rods to form the disordered
structures seen in fig.~\ref{fig:giant} (a)-(d). We should not,
therefore, be surprised that the values of $d$ and $d_{\rm DNA}$,
which we inferred for lamellar structures, deviate from the expected
behavior. The nearly constant behavior of $d_{\rm DNA}$ vs.~$\phi_c$
in the disordered phase can be well observed in the structures seen in
fig.\ref{fig:giant} (a)-(d). The decrease in $d$ in this regime is
also consistent with the transition into the disordered phase, where
the highly charged cationic membranes become squeezed between the
negatively charged DNA rods.

Fig.~\ref{fig:companion} (b) shows the synchrotron X-ray scattering
data reported in ref.~\cite{safinya_science1}. The agreement with the
simulation results in (a) is obvious, lending credibility to our CG
model as well as to the Fourier space analysis of the resulting
structures. The horizontal axes of the figures express the inverse of
the membrane charge density using different scales: $1/\phi_c$ in (a)
and $L/D$ (the mass ratio between the lipid and DNA material) in
(b). The scales are linearly related by: $2.2(1/\phi_c)=L/D$. In both
figures, the lamellar spacing approaches the asymptotic value $d\simeq
2l_{\rm LIP}+D_{\rm DNA}$ at small membrane change densities. These
asymptotic values are different in the two figures, but this is merely
a consequence of the chosen model parameters for $l_{\rm LIP}$ and
$D_{\rm DNA}$, which slightly differ from the experimental
values. Both figures exhibit a weak, and very similar, monotonic
decrease of $d$ with increasing membrane charge density. For fully
charge membranes, the value of $d$ is depressed by about 15-20\%
compared to the low density asymptote. In ref.~\cite{safinya_currop}
this decrease has been attributed to the difference in length between
DOPC (neutral) and DOTAP (cationic), the latter being about 6{\AA}
shorter than the former. We, however, observe the same behavior with
CLs and NLs being geometrically identical. Our simulations point to
two more explanations for this observation: For moderately charged
membranes the effective bilayer thickness slightly shrinks with
$\phi_c$ due to the tensile stress induced by the (negative)
electrostatic energy density of a confined charge neutral systems
\cite{lekner1}. For high membrane charge density, the decrease
in $d$ is likely related to the loss of lamellar order.  In this
regime, the derived value of $d=2\pi/q_{\rm LAM}$ does not necessarily
coincide with the actual interlayer spacing, since the derivation is
based on the presumption that the complex is ideally lamellar.

The agreement between the simulation and experimental results for
$d_{\rm DNA}$ is also clear. In both fig.~\ref{fig:companion} (a) and
(b), we observe that the DNA spacing drops from $d_{\rm DNA}\approx
60${\AA} at low charge densities to $d_{\rm DNA}\approx 30${\AA} at
high charge densities, in a manner which is well approximated by a
linear relationship with the inverse charge density. At very high
charge densities, both figures exhibit the same deviation from a
linear relationship between $d_{\rm DNA}$ and the inverse charge
density. This feature has been attributed in
ref.~\cite{safinya_science1} to the limiting contact distance, $D_{\rm
DNA}$, between DNA rods. This interpretation of the results is correct
provided that the hydration shell is included in the contact
distance. Our results provide yet another possibility. Visual
inspection of the self-assembled structures show that the DNA rods do
not experience any hard core interactions. The plateau-like behavior
of $d_{\rm DNA}$ at high charge densities is related to the formation
of disordered structures which enable a more loose packing of the DNA
rods.

What we have described in this paper is based on the very close
agreement between the computational and experimental results shown in
fig.~\ref{fig:companion}. These results can be explained by a
structural shift from lamellar to fragmented geometry occurring at
high membrane charge densities. The fragmentation of the membranes is
consistent with the membrane rupture observed in our earlier work
described in refs.~\cite{oded_prl,oded_bpj}, where we used a different
CG membrane model. This consistency gives us confidence that the loss
of structural integrity of the membrane at high charge densities is
not an artifact of a particular model. However, we recognize that
although this work features the largest complexes ever simulated,
there may still be some finite size effects that obscure the
comparison with experiments. One such finite size effect is
related to the periodic boundary conditions along the DNA axis, which
enforce the infinite DNA rods to lie parallel to each other. This
constraint, which simplifies the computational scheme, may lead to the
formation of structures with artificial spatial correlations between
the DNA rods. Another finite size effect is related to the relatively
small sizes of the simulated complexes which, therefore, have
scattering plots with peaks that are broader than in the corresponding
experimental scattering plots. This low resolution makes it difficult
to infer the degree of order from the width of the peaks. Yet another
consequence of finite sizes is the enhancement of surface
effects. While it seems plausible that the increase in the
electrostatic tensile stress at high charge densities does proliferate
structural defects, it can be that these defects form more easily on
the boundaries of the complex and, therefore, they become
over-expressed in our smaller complexes. Nevertheless, the close
agreement between fig.~\ref{fig:companion} (a) and (b), over the
entire range of charge densities, supports the possibility that
structural variations observed in our simulations may take place in
nature. Such a structural shift from lamellar to fragmented geometry
should have implications for gene therapy. The shift may explain the
improvement in transfection efficiency (TE) exhibited by these
complexes at high membrane charge densities \cite{transfection1}. One
of the main limiting stages in the transfection process is the release
of the genetic material from the complex into the cytoplasm of the
host cell. It is indeed reasonable to expect that the DNA rods will be
more readily released from the fragmented disordered complexes than
from lamellar structures with long range order. Given the remarkable
success of our model, which is based on a highly CG representation of
the constituting molecular species and their interactions, it is fair
to anticipate the application of this model for obtaining direct
observations of the mechanisms governing transfection and gene
delivery.

\section{Acknowledgments}
We thank Uri Raviv for very useful discussions on X-ray scattering and
Cyrus Safinya for his critical reading of the paper. This work was
supported by the Israel Science Foundation (Grant Number 946/08). NGJ
also acknowledges support from The President's Fund for Visiting Scientists
at Ben Gurion University (Israel) and from Danmarks Nationalbank
(Denmark).

%----------------------------------------------------------- 
%References
%----------------------------------------------------------- 

\newpage

\newpage
  
\newpage

\end{document}